\documentclass{webofc}

\usepackage[varg]{txfonts}   
\usepackage{hyperref}
\usepackage{url}
\hypersetup{colorlinks=true,citecolor=blue,urlcolor=blue,linkcolor=blue}
\usepackage{wrapfig}       
\usepackage{needspace}     
\usepackage{subcaption}    
\RequirePackage{orcidlink}

\newcommand{\sysvar}{\texttt{SysVar}\xspace}
\newcommand{\pyhf}{\texttt{pyhf}\xspace}
\newcommand{\cabinetry}{\texttt{cabinetry}\xspace}

\begin{document}
\title{SysVar}
%
%
\subtitle{Consistent and scalable treatment of systematic uncertainties in template-based analyses}

\author{%
        \firstname{Ilias} \lastname{Tsaklidis}\inst{1}\fnsep\thanks{\email{itsaklid@uni-bonn.de}}\orcidlink{0000-0003-3584-4484} \and
        \firstname{Agrim} \lastname{Aggarwal}\inst{1}\orcidlink{0000-0002-5623-3896} \and
        \firstname{Georgios} \lastname{Alexandris}\inst{1}\orcidlink{0009-0004-3741-2583} \and
        \firstname{Tristan} \lastname{Fillinger}\inst{2}\orcidlink{0000-0001-9795-7412}
        \and
        \firstname{Giacomo} \lastname{De Pietro}\inst{3,4}\orcidlink{0000-0001-8442-107X}     
        \and
        \firstname{Daniel} \lastname{Ivanov}\inst{5}\orcidlink{0000-0002-5423-7009}\and
        \firstname{Melisa-Melek} \lastname{Akdag}\inst{1} \and
        \firstname{Markus} \lastname{Prim}\inst{1}\orcidlink{0000-0002-1407-7450} \and
        \firstname{Florian} \lastname{Bernlochner}\inst{1}\orcidlink{0000-0001-8153-2719}
}

\institute{Physikalisches Institut, University of Bonn, 53115 Bonn, Germany
  \and High Energy Accelerator Research Organization (KEK), Tsukuba 305-0801, Japan 
  \and Insitute for Experimental Particle Physics (ETP), Karlsruhe Institute of Technology (KIT), Karlsruhe, Germany
  \and Scientific Computing Center (SCC), Karlsruhe Institute of Technology (KIT), Karlsruhe, Germany
  \and Graduate School of Science, Nagoya University, Nagoya 464-8601, Japan
          }

\abstract{
  Propagating systematic uncertainties from correction weights into templates generated from simulated data while preserving correlations in the signal extraction variables becomes increasingly challenging as analyses scale in size.
  SysVar's lightweight API hides most of the bookkeeping complexity of applying correction weights, building systematic variations, and histogramming nominal and varied templates.
  This streamlines workflows that are otherwise error-prone but essential for high-precision measurements.
  Correlations are preserved by propagating the full covariance across the analysis and compressing it into a small set of orthogonal eigenvariations, each defining an ordinary shape nuisance parameter.
  In this work, we demonstrate \sysvar's workflow and impact by combining two independent pseudo-measurements that benefit from consistently correlated systematic uncertainties encoded in the template shapes.
  SysVar is open-source, pip-installable, and experiment-agnostic.
}
\maketitle

\section{Introduction}
\label{intro}

In a typical high-energy physics measurement, the observed data is a mixture
of several physical processes, isolated through a template
fit~\cite{bakercousins} against simulation. This simulation is reweighted to match data,
through efficiency, resolution or physics-motivated corrections and such reweighting introduces systematic
uncertainties. These corrections are typically measured in bins of auxiliary
variables, such as particle momentum or angle. Whatever correlations exist
between their bins must be preserved once the corresponding uncertainties
reach the templates, where they are implemented as nuisance parameters.

Frameworks built on the HistFactory model~\cite{histfactory}, such as
\pyhf~\cite{pyhf,pyhf_joss}, support nuisance parameters that are fully correlated or fully uncorrelated
across the bins of the signal-extraction variable, but nothing in between.
Reproducing a partial correlation therefore requires decomposing it by hand
into nuisance parameters the framework supports. This means tracking
which correction acts on which bins and combining correlated sources into a
single covariance, which becomes impractical once the number of corrections,
channels and templates grows. The common approximation is instead to treat
such correlations as either full or absent, since that is what these
frameworks support directly. This generally does not hold as we show in
Section~\ref{dataset}.

\sysvar addresses this by sampling correlated variations of the correction
weights, building the resulting covariance across the full analysis, and
compressing it through an eigendecomposition into a small set of orthogonal
eigenvariations. Each eigenvariation defines an ordinary, fully correlated
shape nuisance parameter that a standard fitting framework can use, with
native output for \cabinetry~\cite{cabinetry} and \pyhf. Analyses are defined
declaratively rather than through procedural code, and every intermediate
step of the propagation remains inspectable. Although developed for
Belle~II, nothing about the package is experiment-specific.

The rest of this document is organized as follows. We introduce a synthetic
dataset illustrating why this propagation is non-trivial in
Section~\ref{dataset}. We describe the method itself in Section~\ref{method}.
We demonstrate its impact on a toy combination of two measurements in
Section~\ref{combinations}. We discuss its usage and computational cost in
Sections~\ref{usage} and~\ref{performance}, and conclude in
Section~\ref{conclusion}.

\section{Dataset and problem}
\label{dataset}

For demonstration purposes, we generate a synthetic dataset representing a
generic decay $X \rightarrow YZ$.

\needspace{9cm}
\begin{wrapfigure}{r}{0.46\textwidth}
  \centering
  \includegraphics[width=0.44\textwidth]{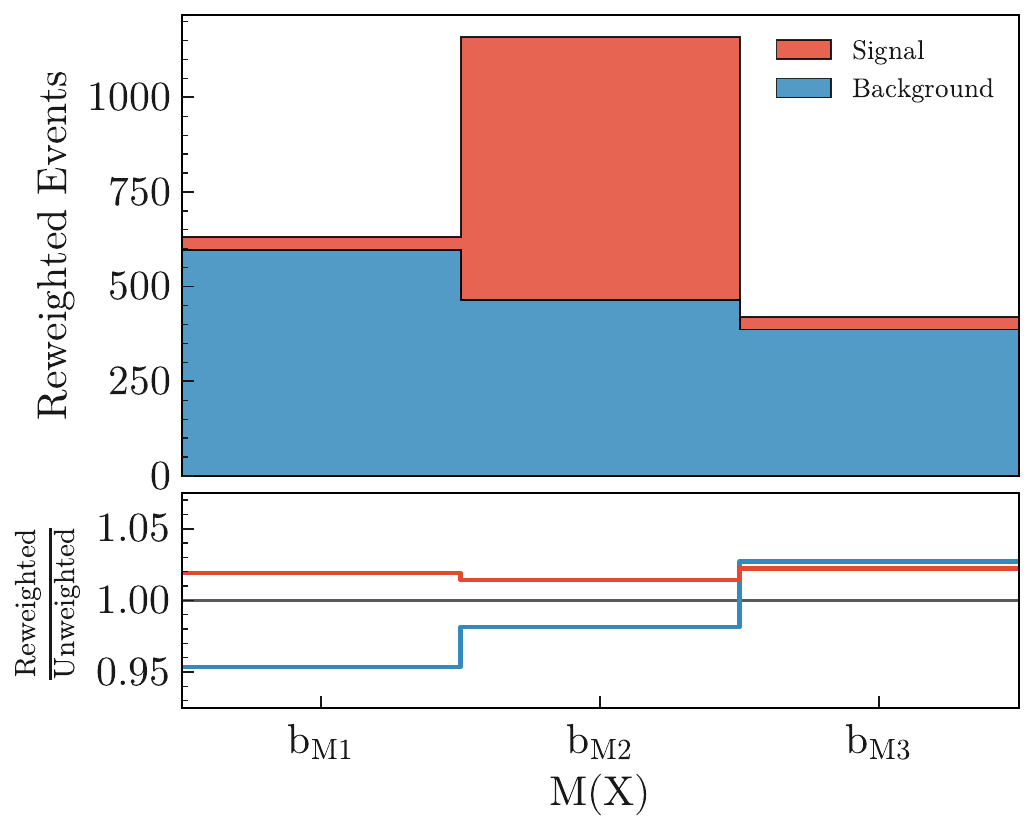}
  \caption{Stacked signal and background templates in the fit observable $M$,
  after applying the calibration weights (top), and the ratio of the yield
  with and without the calibration weight in each bin, shown separately for
  the two templates (bottom).}
  \label{fig:templates}
\end{wrapfigure}

Each event is labelled as either signal or background and carries two
quantities. The fit observable $M$ is the invariant mass of the $X$ system.
The calibration variable $c$ stands for the quantity in which an auxiliary
calibration is measured, for instance the momentum of the decay product $Y$.
This is deliberately not the observable used in the fit. We denote the
$i^{\mathrm{th}}$ bin of an observable $O$ by $b_{O_i}$, and its partial
contribution from template $t$ by $b^{t}_{O_i}$.

Whether an event is signal or background depends on $c$, but only weakly. The
dependence is far too weak to separate the two processes in a fit. It instead
stands in for the hidden relations between the composition of a sample and
the variables in which its corrections are binned, relations that are in
practice rarely understood in detail. Generating the labels independently of
$c$ would be the unrealistic choice. The correction would then factor out of
the fit observable, and the effect under study would disappear by
construction.


The correction under study is an efficiency calibration measured in two bins of
$c$, denoted $b_{C_1}$ and $b_{C_2}$, with per-bin uncertainties at the percent
level. Every event is reweighted
by the calibration weight of the bin it falls into. Signal and background are
given different shapes in $M$, which is what allows them to be separated in the
fit. Figure~\ref{fig:templates} shows the effect. The correction
  shifts the normalisation of both templates and changes their shape
  differently for signal and background, because the two processes populate
  the calibration bins differently.

So far this is only a statement about central values. The actual difficulty
is that the calibration's own uncertainties are, in general, partially
correlated: they are derived from auxiliary measurements, and even
statistically independent control samples share backgrounds and systematic
uncertainties. The auxiliary measurement therefore delivers a covariance,
parametrised here by $\rho_{C} = \rho(b_{C_1}, b_{C_2})$, for which we take an
illustrative value of $0.1$.

\begin{figure}[h!]
  \centering
  \begin{subfigure}[b]{0.345\textwidth}
    \centering
    \includegraphics[width=\textwidth]{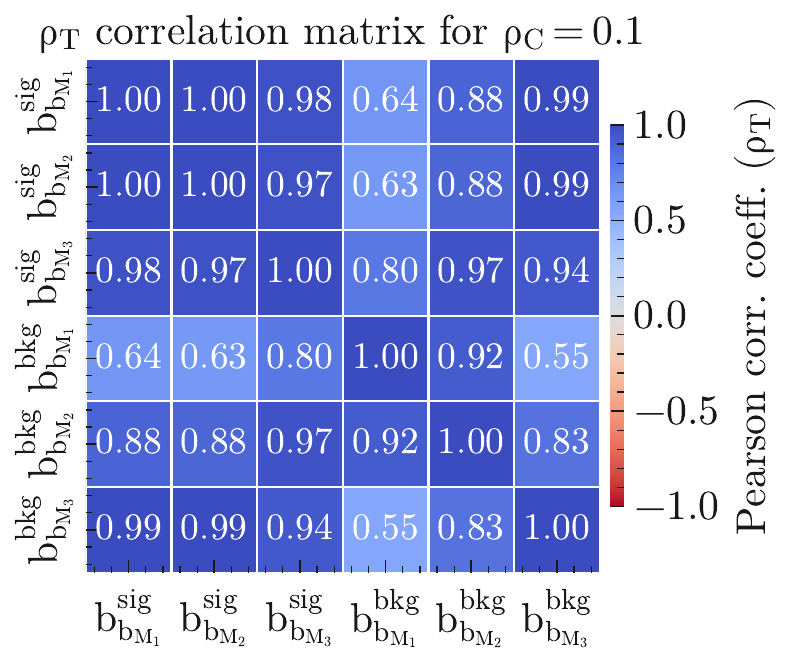}
    \caption{}
    \label{fig:tmplcorr}
  \end{subfigure}
  \hfill
  \begin{subfigure}[b]{0.595\textwidth}
    \centering
    \includegraphics[width=\textwidth]{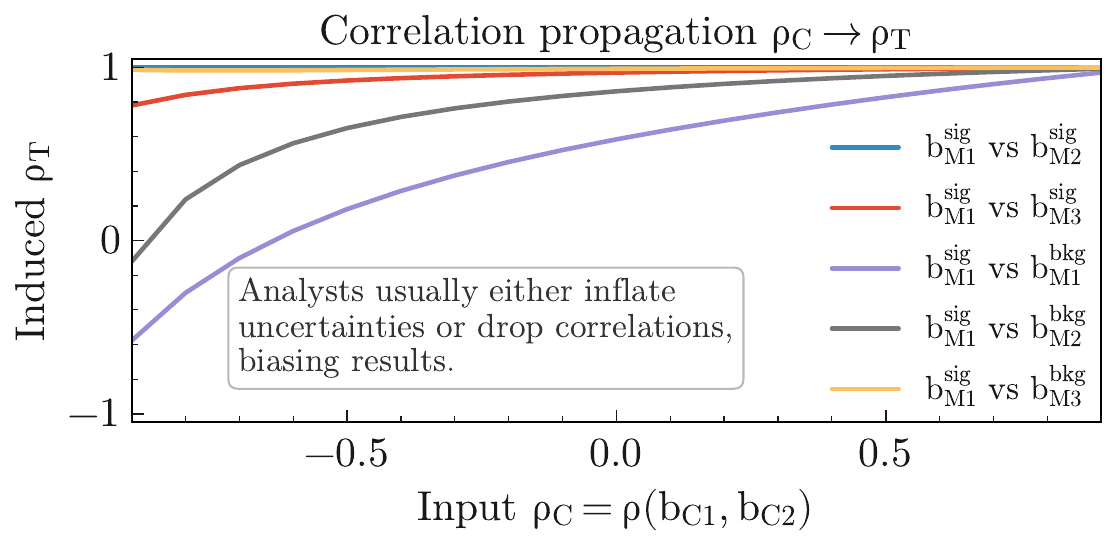}
    \caption{}
    \label{fig:scan}
  \end{subfigure}
  \caption{(a) Correlation matrix $\rho_{T}$ induced across the six bins of the
  analysis (three bins of $M$ for each of the two templates) by a
  correction whose two calibration bins are correlated with $\rho_{C} = 0.1$.
  (b) Induced correlation $\rho_{T}$ between the first signal bin and each of
  the remaining five bins, as a function of $\rho_{C}$, with all other inputs
  held fixed.}
  \label{fig:corr}
\end{figure}

This covariance does not stay in the space of the calibration once
propagated through the reweighting. Each template bin receives contributions
from both calibration bins, so $\rho_{C}$ re-emerges as a correlation between
bins of the fit observable, crossing templates and template types alike. The
result is the dense $6 \times 6$ matrix $\rho_{T}(b^{t_p}_{M_i},
b^{t_q}_{M_j})$ of Figure~\ref{fig:tmplcorr}, with
$t \in \{\mathrm{sig}, \mathrm{bkg}\}$.

This relation is neither linear nor universal. Figure~\ref{fig:scan} scans
$\rho_{T}$ between the first signal bin and each of the other five across the
full range of $\rho_{C}$. The pairs respond very differently, from nearly
insensitive to swinging between $-0.6$ and $+1$, and all remain large and
positive already at $\rho_{C} = 0$, so treating them as uncorrelated is not
conservative but wrong.

\section{Method for propagating the correction uncertainties into templates}
\label{method}

The correlations described in Section~\ref{dataset} cannot be written in
closed form. They depend on how the events of each template distribute
across the bins of every correction. \sysvar instead propagates them by
sampling. The propagation happens in four stages. \sysvar builds the
covariance of the correction weights, draws correlated variations from it,
rebuilds the templates for each variation, and turns the resulting
covariance into a basis a fitting framework can use.

\paragraph{Sampling.}
All uncertainty sources acting on a correction are collected into a single
covariance matrix, and $N_v$ variations, or toys, are drawn from the
multivariate Gaussian it defines, reproducing any correlation the
auxiliary measurement reported.
If several objects in an event receive a weight from
the same correction (say, two leptons with the same identification
efficiency) in the same bin, \sysvar applies the same toy's values to all
of them. Varying each independently would treat the correction bins as
independent within the event, defeating the purpose of the correlated draw.

\paragraph{Building the covariance.}
Each variation gives a re-histogrammed set of templates. Let $N_{vI}$ be the
content of bin $I$ in variation $v$, with composite index $I = (c, s, i)$
labelling bin $i$ of template $s$ in channel $c$. Let
$\Gamma_{vI} = N_{vI} - \bar{N}_{I}$ be its shift from the mean over
variations. The covariance of the systematic source over the whole analysis is

\begin{equation}
  C_{\mathrm{syst}}^{IJ} = \frac{1}{N_v - 1} \sum_{v=1}^{N_v}
    \Gamma_{vI}\, \Gamma_{vJ} .
  \label{eq:cov}
\end{equation}

$I$ and $J$ range over channels, templates and bins together, so this is a
single covariance over the full analysis space rather than one matrix per
template. It is shown, in normalised form, in Figure~\ref{fig:tmplcorr}.

\paragraph{Change of basis.}
A covariance is not something a template fit can use directly. Frameworks
built on the HistFactory model, such as \pyhf, represent
a systematic as a nuisance parameter whose bin-to-bin behaviour is either
fully correlated, with one parameter driving every bin together, or fully
uncorrelated, with one parameter per bin, and nothing in between.
Diagonalising the covariance,

\begin{equation}
  C_{\mathrm{syst}} = V U V^{T} = \left(V\sqrt{U}\right)
    \left(V\sqrt{U}\right)^{T} = \Gamma' \Gamma'^{T} ,
  \qquad \Gamma' = V\sqrt{U} ,
  \label{eq:eigen}
\end{equation}

where $V$ is the matrix of eigenvectors of $C_{\mathrm{syst}}$ and $U$ the
diagonal matrix of its eigenvalues, turns each eigenvector into an eigenvariation. Each eigenvariation is a
single coherent, orthogonal shift of every bin that sets the size of an
ordinary, fully correlated shape nuisance parameter's effect. Retaining every
eigenvariation reproduces $C_{\mathrm{syst}}$ exactly, so nothing is
approximated at this stage.

\begin{figure}[h!]
  \centering
  \includegraphics[width=\textwidth]{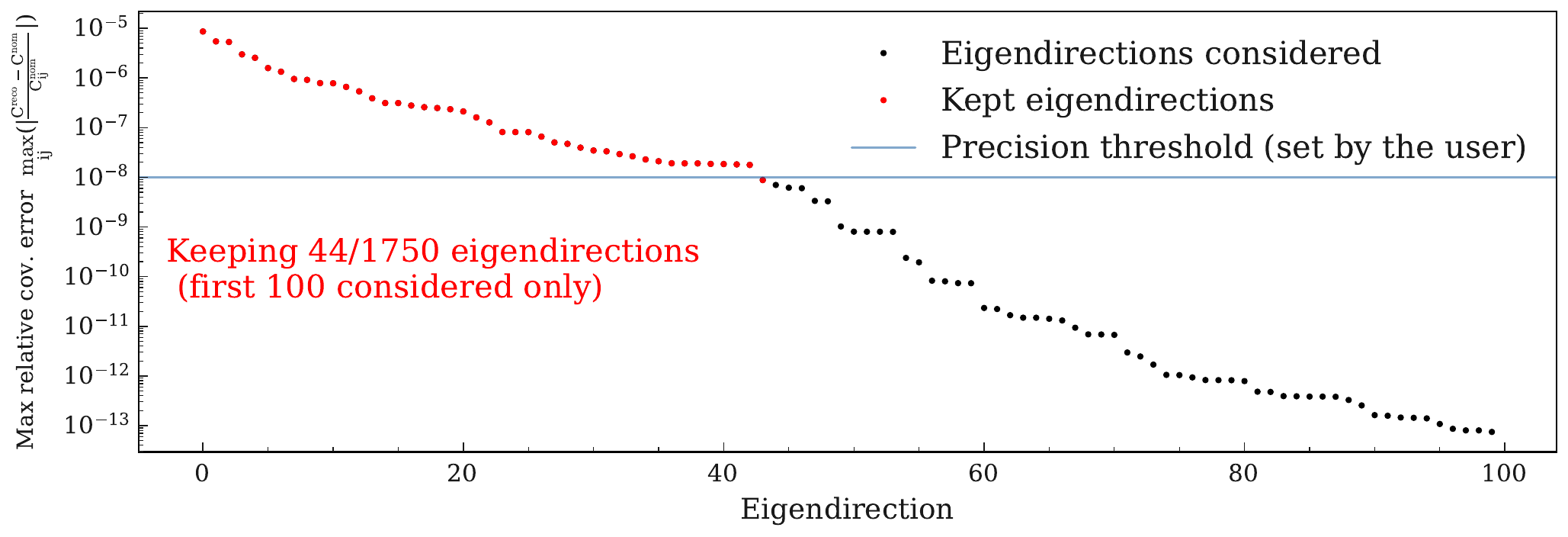}
  \caption{Convergence of the truncation for a large analysis with $1750$
  template-space bins, as eigenvariations are added in decreasing eigenvalue
  order. The largest residual entry of $C_{\mathrm{syst}} - C'_{\mathrm{syst}}$,
  normalised to the largest entry of $C_{\mathrm{syst}}$, drops below the
  $10^{-8}$ precision threshold after $44$ eigendirections.}
  \label{fig:truncation}
\end{figure}

\paragraph{Truncation.}
Without truncation, the number of shape nuisance parameters would equal the
dimension of the template space, one per bin. This is rarely necessary, because
the eigenvalues of $C_{\mathrm{syst}}$ typically fall off by orders of magnitude
within the first few dozen directions, well short of that full dimension, so a
small subset already reproduces the covariance to a chosen precision. \sysvar
adds eigenvariations in order of decreasing eigenvalue and stops once the
largest residual entry of $C_{\mathrm{syst}} - C'_{\mathrm{syst}}$, normalised to
the largest entry of $C_{\mathrm{syst}}$, drops below a user-set threshold. A
criterion based on the accumulated trace is available as an alternative.
Figure~\ref{fig:truncation} shows this convergence for a large analysis.

This criterion is a statement about the covariance rather than the measurement,
and is best read as a starting point rather than a guarantee. It should be
validated case by case. Add the retained eigenvariations to the fit one at a
time until the uncertainty on the parameter of interest stops moving. If it has
not stopped moving by the last saved one, the requested precision was too loose
and the eigendecomposition should be repeated with a tighter one.

How many directions are retained is itself informative. A fully correlated
correction collapses onto one dominant eigendirection, an uncorrelated one
onto one per weight, and the partially correlated case in between, the one
that motivates \sysvar, needs a number that cannot be anticipated. For the
dataset of Section~\ref{dataset}, whose correction has only two calibration
bins, the covariance has rank two, and \sysvar retains exactly two
eigendirections.

The retained eigenvariations are written out as the up and down template
shifts that \cabinetry~\cite{cabinetry} and \pyhf expect for a shape nuisance
parameter, so the output feeds directly into the signal-extraction fit.

\section{Impact on measurement combinations}
\label{combinations}

The case where a consistent treatment matters most is the one where a systematic
uncertainty is shared between measurements that are otherwise independent. To
illustrate it we built a toy combination along the lines of Ref.~\cite{marco},
in which Belle~II and LHCb pseudo-measurements of the lepton flavour
universality ratios
$R(D^{(*)}) = \mathcal{B}(B \to D^{(*)}\tau\nu) / \mathcal{B}(B \to D^{(*)}\ell\nu)$ are brought together.

The setup uses four reconstruction channels, an $R(D)$ and an $R(D^{*})$
channel for each experiment, built on the fit model of Ref.~\cite{marco}. For
details on the signal-extraction variables and templates we refer the reader
to that reference. That study does not include systematic uncertainties, so
for this demonstration we introduce a single shared source of our own. Three
background branching-fraction corrections, $\mathrm{BF}_1$, $\mathrm{BF}_2$
and $\mathrm{BF}_3$, are assumed to be partially correlated, as shown in
Table~\ref{tab:bfcorr}.

\begin{wraptable}{r}{4.4cm}
  \centering
  \begin{tabular}{c|ccc}
    & $\mathrm{BF}_1$ & $\mathrm{BF}_2$ & $\mathrm{BF}_3$ \\
    \hline
    $\mathrm{BF}_1$ & $1$     & $0.05$  & $-0.35$ \\
    $\mathrm{BF}_2$ & $0.05$  & $1$     & $0.10$  \\
    $\mathrm{BF}_3$ & $-0.35$ & $0.10$  & $1$     \\
  \end{tabular}
  \caption{Correlation assumed between the three background
  branching-fraction corrections.}
  \label{tab:bfcorr}
\end{wraptable}

Two ways of arriving at a combined result are then compared. The conventional
one fits each experiment separately and averages the outcomes afterwards, with
the shared systematic declared fully correlated between them. This is an
assumption that can only be applied to the final uncertainties, since neither
fit knows anything about the other's templates. The alternative is a single
fit over all four channels in which the shared uncertainty enters through the
eigenvariations built from the common covariance, so that it acts on the
shapes of the templates in both experiments at once, and its correlations
are carried through the fit rather than imposed on the results.

\begin{figure}[h!]
  \centering
  \includegraphics[width=0.66\textwidth]{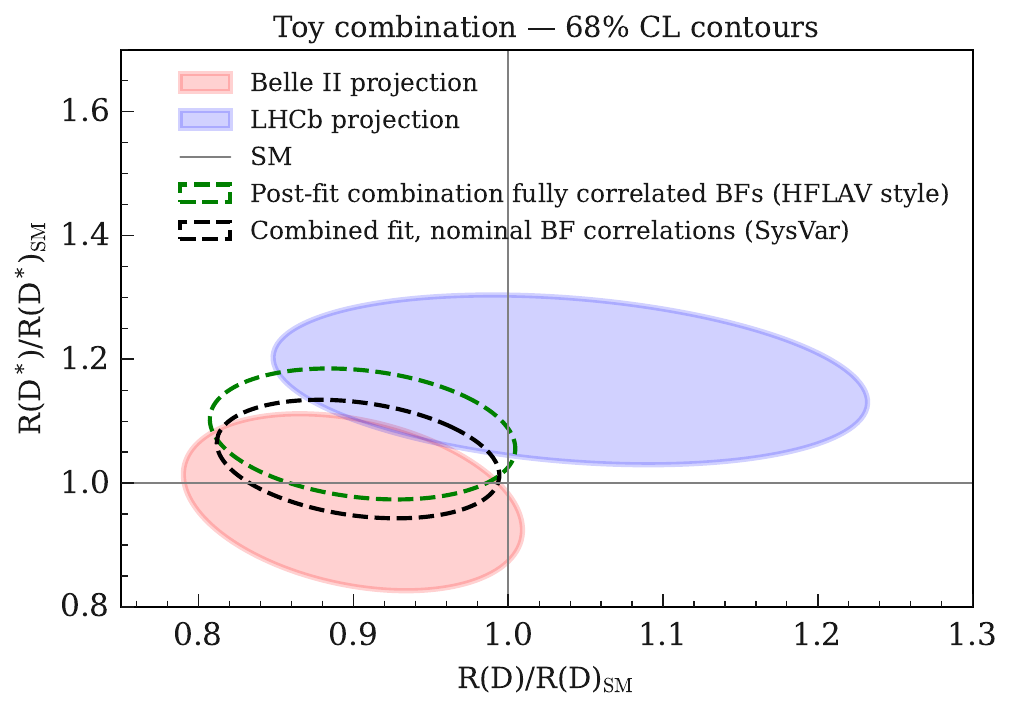}
  \caption{Toy combination of Belle~II and LHCb pseudo-measurements of $R(D)$
  and $R(D^{*})$, normalised to their Standard Model values. The filled contours
  are the two individual pseudo-measurements. The dashed contours compare a
  post-fit average in which the shared branching-fraction uncertainty is assumed
  to be fully correlated between the experiments with a simultaneous fit in
  which the same uncertainty is propagated into the template shapes by \sysvar.
  All contours are at $68\%$ confidence level. Relative to the post-fit
  average, the combined fit's central value differs by about $-4\%$ in
  $R(D^{*})$ (roughly $0.4\,\sigma$) and by less than $0.5\%$ in $R(D)$
  (about $0.03\,\sigma$).}
  \label{fig:combination}
\end{figure}

Figure~\ref{fig:combination} shows the two. The contours are of comparable
size, but they are not centred on the same point. The combined fit sits
lower in $R(D^{*})$ than the post-fit average. The reason is that a
correlation asserted between two final uncertainties acts only on their
sizes, whereas propagating the shared source through the templates lets it
act on their shapes. Where a common background is pulled in one direction by
the data of one experiment, the correlated shape uncertainty transmits that
constraint to the other, and the signal yields respond accordingly.

This comparison is illustrative rather than predictive. It comes from a
single pseudo-dataset, so the exact size and direction of the shift belong
to that realisation. What it does show is that the two treatments are not
equivalent. Once experiments share a systematic source, the choice of how to
correlate it is a modelling decision that moves the combined central value,
and propagating the source through the template shapes replaces a
correlation imposed afterward with one that the fit itself can act on.

\section{Usage}
\label{usage}

Everything described in Section~\ref{method} is reached through two function calls.
A helper function  attaches the nominal correction weights,
read from a \texttt{CSV} file or a Python dictionary, to the events.
Another one then samples the variations, rebuilds the templates,
forms the covariance of Equation~\ref{eq:cov} and returns the eigenvariations.
The sampling, histogramming and bookkeeping are all handled internally.

Convenience like this can come at the price of opacity, which \sysvar avoids by
construction. It is organised by class composition with only shallow
inheritance. Each stage, from the uncertainty models and corrections through
the variations and templates to the eigendecomposition, is a self-contained
object held by the next stage rather than derived from it. The object
returned by \texttt{eigendecompose} keeps every intermediate reachable, from
the covariance and eigenvariations down to the individual varied templates,
each with its own plotting methods and complemented by further visualisation
entry points in the API. \sysvar streamlines the workflow without hiding it.
Every step can be inspected.

What the analyst writes is a declaration, not code. Channels, templates and
systematic sources are entries in a single configuration object, in Python,
\texttt{YAML} or \texttt{JSON}, that lives in one place, can be
version-controlled and diffed like any other input, and grows by adding
entries rather than writing procedural code. A correction can be included in or
excluded from specific channels or templates just as easily, with the
resulting combinatorics resolved by \sysvar rather than by hand. The retained
eigenvariations are written to a \texttt{ROOT}~\cite{root} file laid out as \cabinetry
expects, so the transition to the fit needs no further bookkeeping.

\section{Performance and scaling}
\label{performance}

Whether this is usable in practice depends on what it costs to run. We measured
this with a standalone benchmark that generates a synthetic sample like that of
Section~\ref{dataset} and times only the \sysvar propagation of
Section~\ref{method}, scanning a four-dimensional space around a fixed
baseline, made up of the size of the event sample, the number of correction
bins, the binning of the fit observable, and the number of objects per event
that share a correction. Each configuration was repeated $20$ times on a
single process, recording wall-clock time and peak resident memory.

\begin{figure}[h!]
  \centering
  \includegraphics[width=0.95\textwidth]{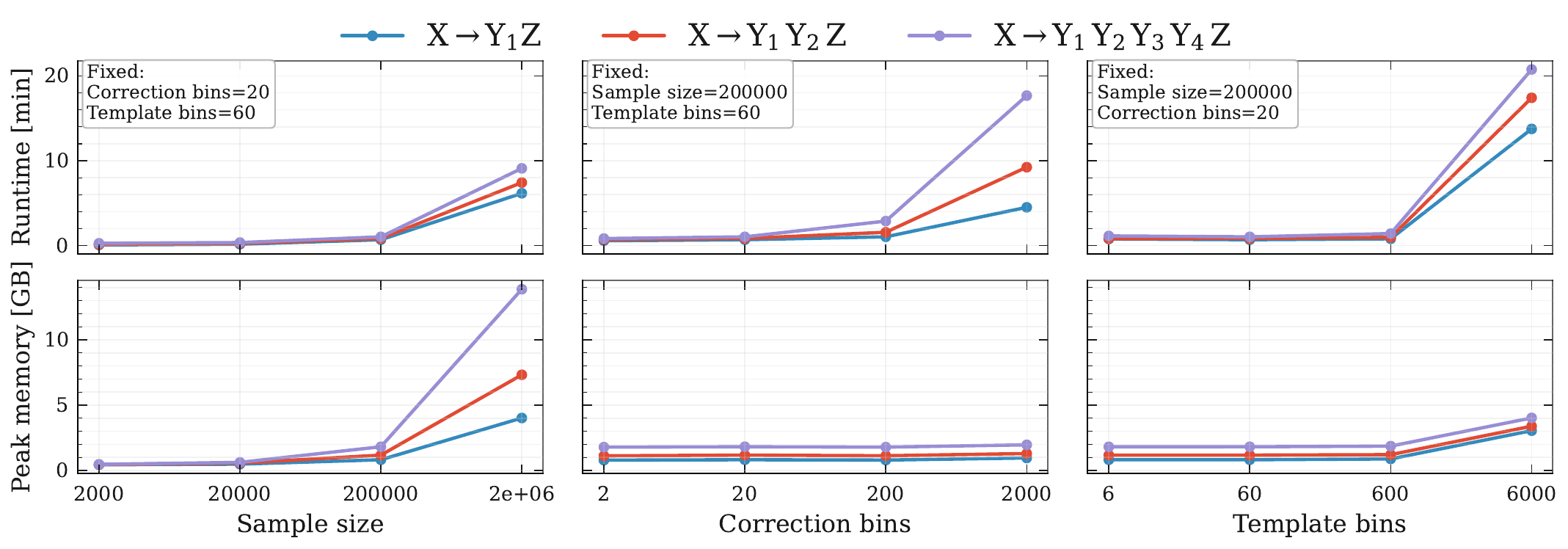}
  \caption{Runtime (top) and peak memory (bottom) of the \sysvar stage as each
  scanned quantity is varied around the baseline, for one, two and four objects
  per event carrying a weight from the same correction.}
  \label{fig:performance}
\end{figure}

Figure~\ref{fig:performance} shows the resulting means. Memory scales as
expected for a dataframe-based tool, growing linearly with the event sample and
the number of corrected objects per event. Runtime is mild elsewhere but rises
steeply with the binning of the fit observable, since diagonalising the
covariance of Equation~\ref{eq:cov} costs $\mathcal{O}(d^3)$, where $d$ is the
dimension of the template space~\cite{golubvanloan}. For typical sample sizes and binnings the
propagation costs minutes and a few gigabytes on a single core. What
eventually limits the method is not the size of the dataset but the
dimension of the template space, and the truncation of Section~\ref{method}
keeps that from propagating into the number of nuisance parameters carried
by the fit.

\section{Conclusion and outlook}
\label{conclusion}

We have presented \sysvar, a package for propagating the uncertainties of
simulation correction weights into template-based analyses while preserving
their correlations. Using a synthetic dataset, we demonstrated how a
correction with partially correlated per-bin uncertainties induces a dense
correlation structure across the fit templates that neither a fully
correlated nor an uncorrelated treatment reproduces. We described how
\sysvar builds the corresponding covariance and compresses it into a small
set of orthogonal eigenvariations, each defining an ordinary nuisance
parameter. We showed its impact on a toy combination of two measurements
sharing a systematic source, where propagating the shared uncertainty
through the template shapes shifted the combined result relative to a
conventional post-fit average. The propagation itself runs in minutes on a
single core for realistic analyses.

The most significant item on the roadmap is a change to where the
eigendecomposition itself takes place. \sysvar currently builds the covariance
from the varied templates and diagonalises it in the, typically large,
template space. A mathematically equivalent alternative is to diagonalise the
covariance of the correction weights themselves (a much smaller matrix) and
vary the templates directly along its orthogonal eigendirections. This has
not been implemented yet. In developing \sysvar, corrections were
deliberately left untransformed at first, so that every intermediate
quantity could be compared one to one against existing, template-level
analyses. Having now demonstrated that \sysvar reproduces the expected
behaviour, attaching the central values of the correction weights and
diagonalising the corrections directly is the more straightforward and
intuitive approach. Supporting this correction-space eigendecomposition as an
additional option is the main development planned, and would in particular
reduce the cost of the diagonalisation discussed in Section~\ref{performance}.

Beyond this, the API will keep growing to make analyses easier to declare,
and its visualisation methods will keep extending. Further performance work
will target runtime and memory usage, in particular through parallelisation,
which is a promising route for the largest analyses.

\sysvar already has a growing base of users within Belle~II, where the
performance group is encouraging its adoption and has aligned its recommended
format for correction weights with the one \sysvar expects. Nothing about it,
however, is specific to Belle~II. Its features are experiment-agnostic,
making it equally suited to an external user, or to anyone with access to
tuples from more than one experiment who wants to combine them consistently.

\bibliography{references}

\end{document}